\documentclass[smallextended,referee]{svjour3}

\usepackage{algorithm}
\usepackage{algorithmic}
\usepackage{amsmath}
\usepackage{amssymb}
\usepackage{graphicx}
\usepackage{multirow}
\usepackage{subfigure}
\usepackage{url}
\usepackage{colortbl}
\usepackage[numbers]{natbib}

\newcolumntype{V}{>{$\vcenter\bgroup\hbox\bgroup}c<{\egroup\egroup$}}

\newcommand{\topincludegraphics}[2][]{%
  \raisebox{\dimexpr-\height+\ht\strutbox\relax}{\includegraphics[#1]{#2}}}

\begin{document}

\title{A Note on the Capability Profile -- Localized Increase in Force Production and Its Effect on Overall Resistance Training Performance}
\titlerunning{Localized Increase in Force Production and its Effect on Performance}

\author{Ognjen Arandjelovi\'c}

\institute{
   Centre for Pattern Recognition and Data Analytics\\[-5pt]
   Deakin University\\[-5pt]
   Australia\\[-5pt]
   E-mail: \texttt{ognjen.arandjelovic@gmail.com}
}

\date{}

\maketitle

\begin{abstract}
In this article I resolve formally the apparent paradox that arises in the context of the computational model of resistance exercise introduced in my previous work. Contrary to intuition, the model seems to allow for a localized increase in force production to degrade the ultimate exercise performance. I show this not to be the case.
\end{abstract}

\section*{Resolution of the apparent paradox}
Following the publication of my article in which the concept of an exercise capability profile of an athlete was first introduced~\cite{Aran2010-med} and subsequent studies which adopted it as the basis for further inquiry~\cite{Aran2011-med,Aran2012-med,Aran2013a-med,Aran2013d}, I was contacted by a number of researchers and strength practitioners who sought further insight regarding a particular observation made therein. I cite the key paragraph from~\cite{Aran2011-med}:
\begin{quotation}
``Determining the optimality of a training approach is further complicated by the dependence of force produced by muscle on its elongation and rate of contraction. Alteration of force produced at a certain point in a lift affects the velocity of the bar, and with it the rate of contraction of muscles involved, throughout the remainder of the lift. Higher velocity of the bar can then offset, or more than offset, lower force production at a position of increased velocity in a lift. Thus and seemingly paradoxically, even if force production characteristics of the athlete are only increased, the overall performance may not be improved.''
\end{quotation}
The question which interested many readers who got in touch with me is if an increase in the velocity of the bar can affect subsequent force production to such a degree so as to result in a \emph{decrease} of the overall performance, i.e.\ the load used in an exercise. The practical significance of this question is clear and the methodology which was adopted in the original paper can be used to reach the answer.

\begin{figure}[htb]
  \vspace{20pt}
  \centering
  \footnotesize
  \begin{tabular}{ccc}
    \topincludegraphics[width=0.4\textwidth]{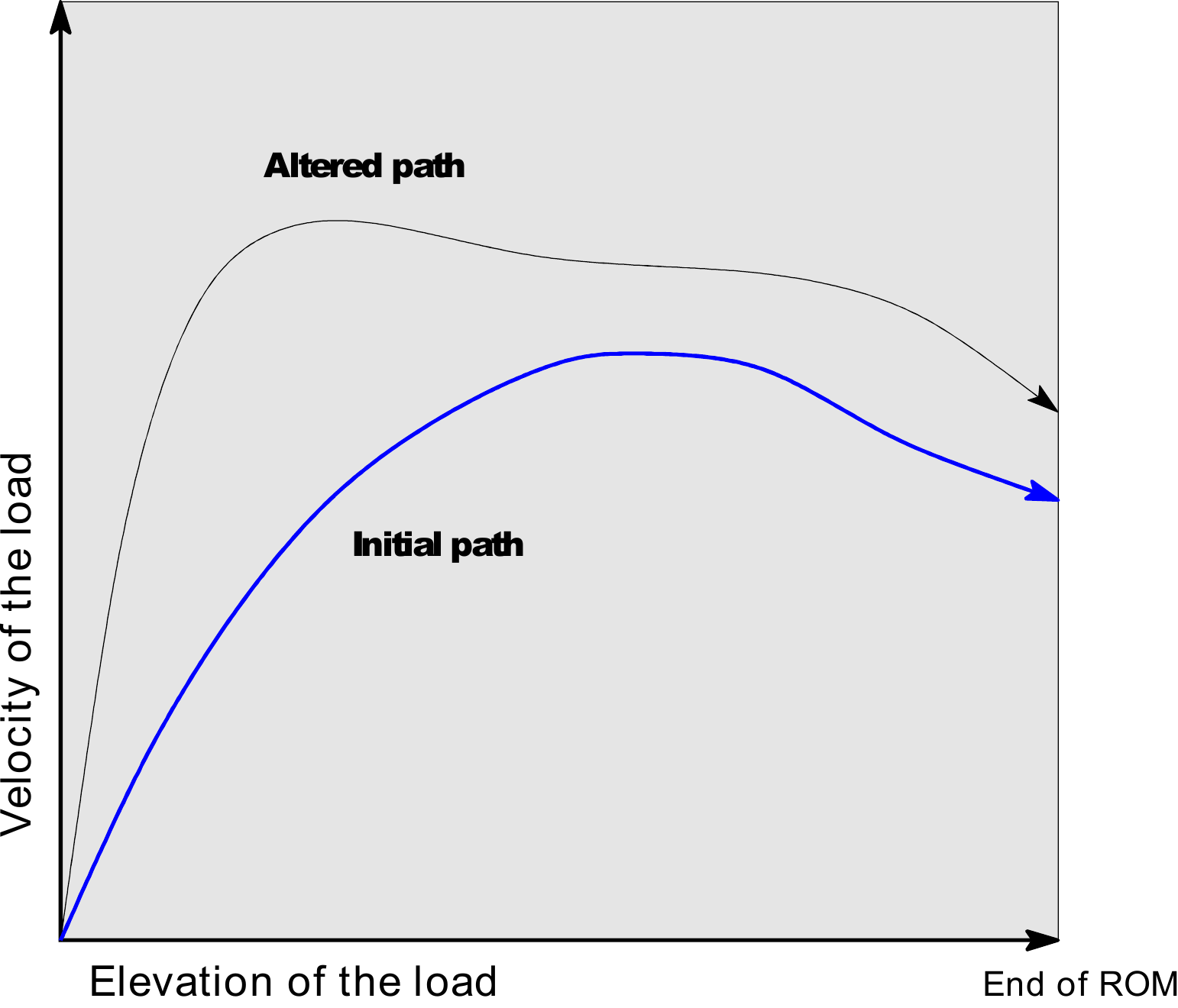} &~\hspace{10pt}~& \topincludegraphics[width=0.4\textwidth]{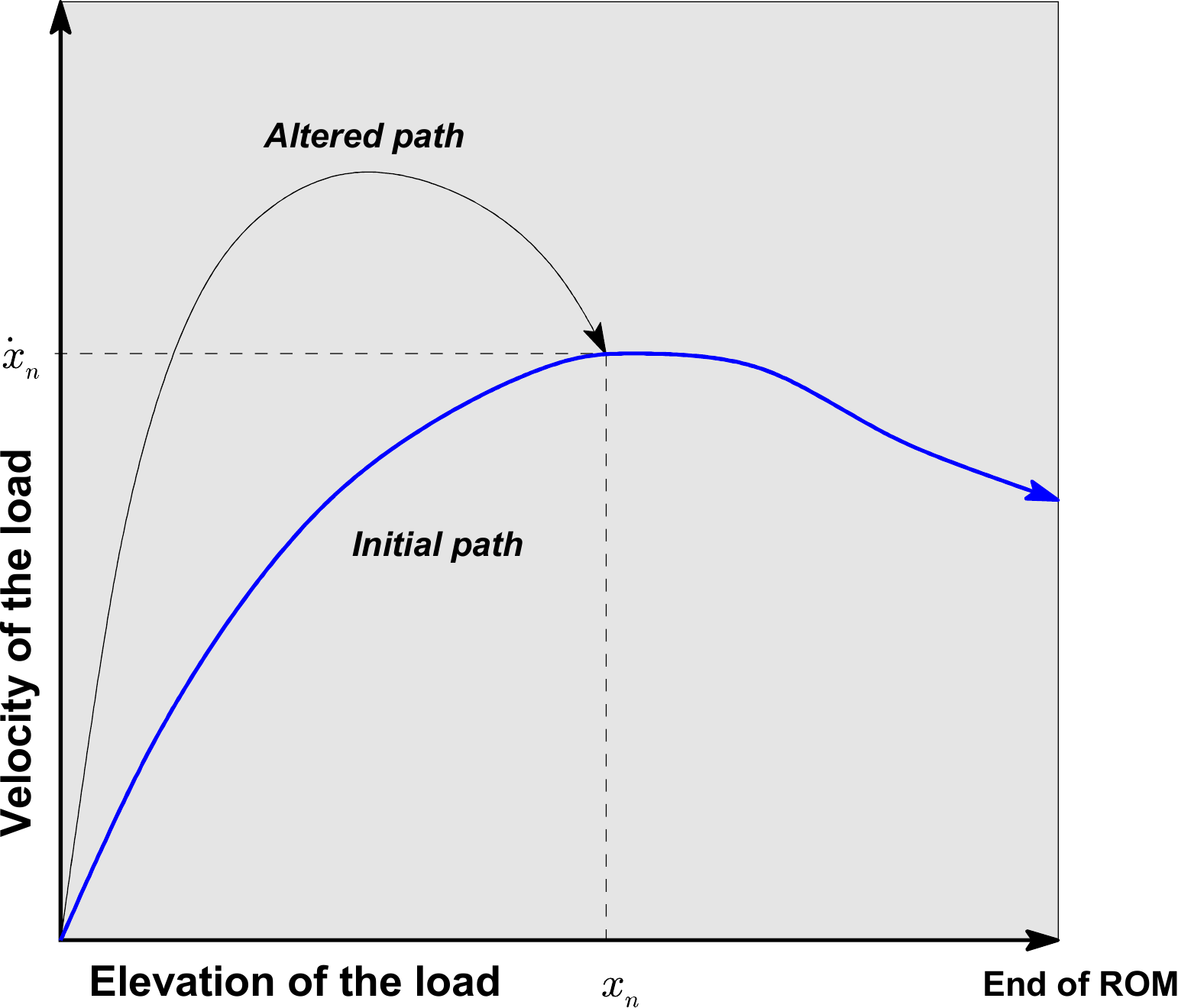}\\
    (a) && (b)\\
    &&\\[-5pt]
    \topincludegraphics[width=0.4\textwidth]{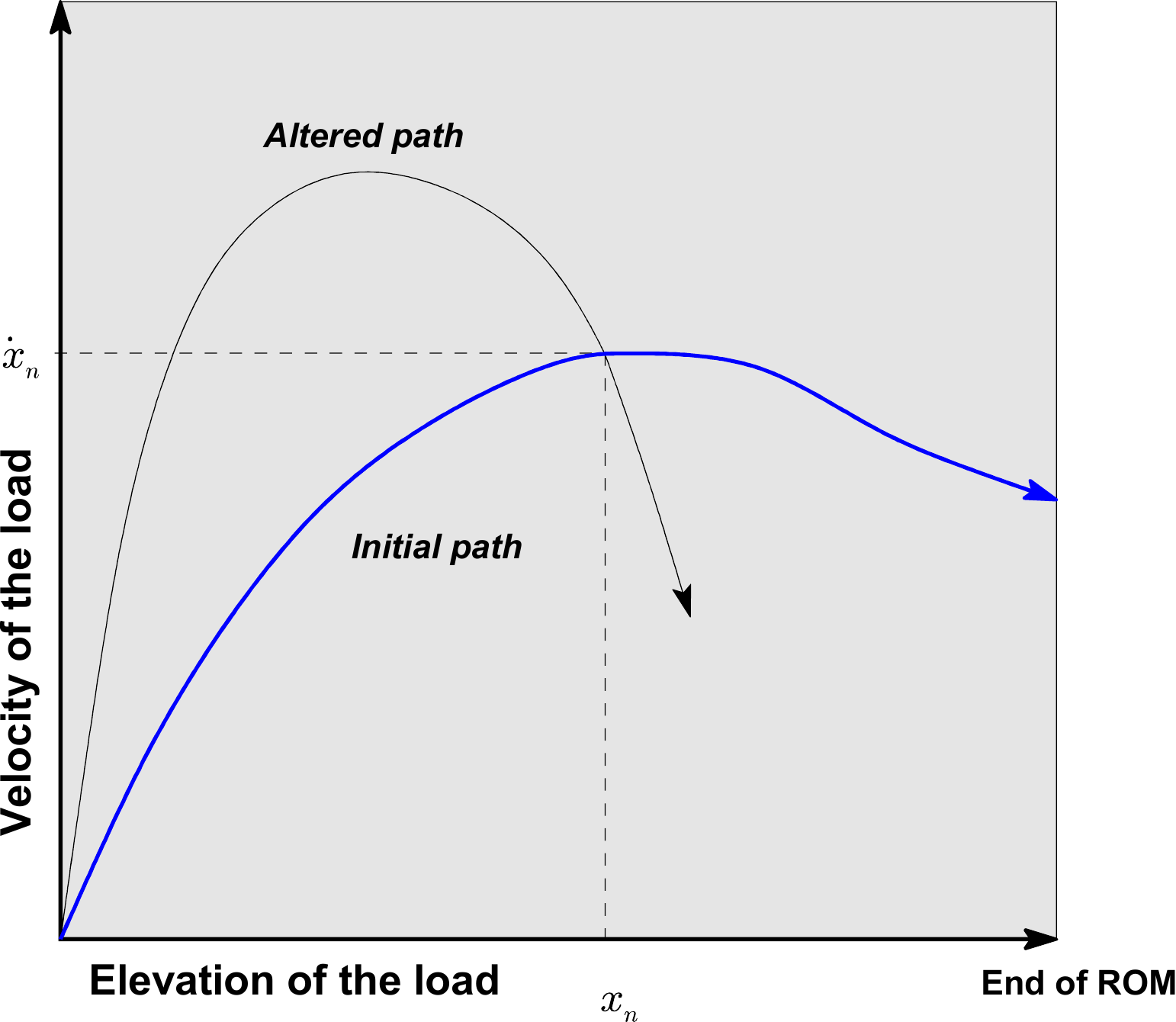} && \topincludegraphics[width=0.4\textwidth]{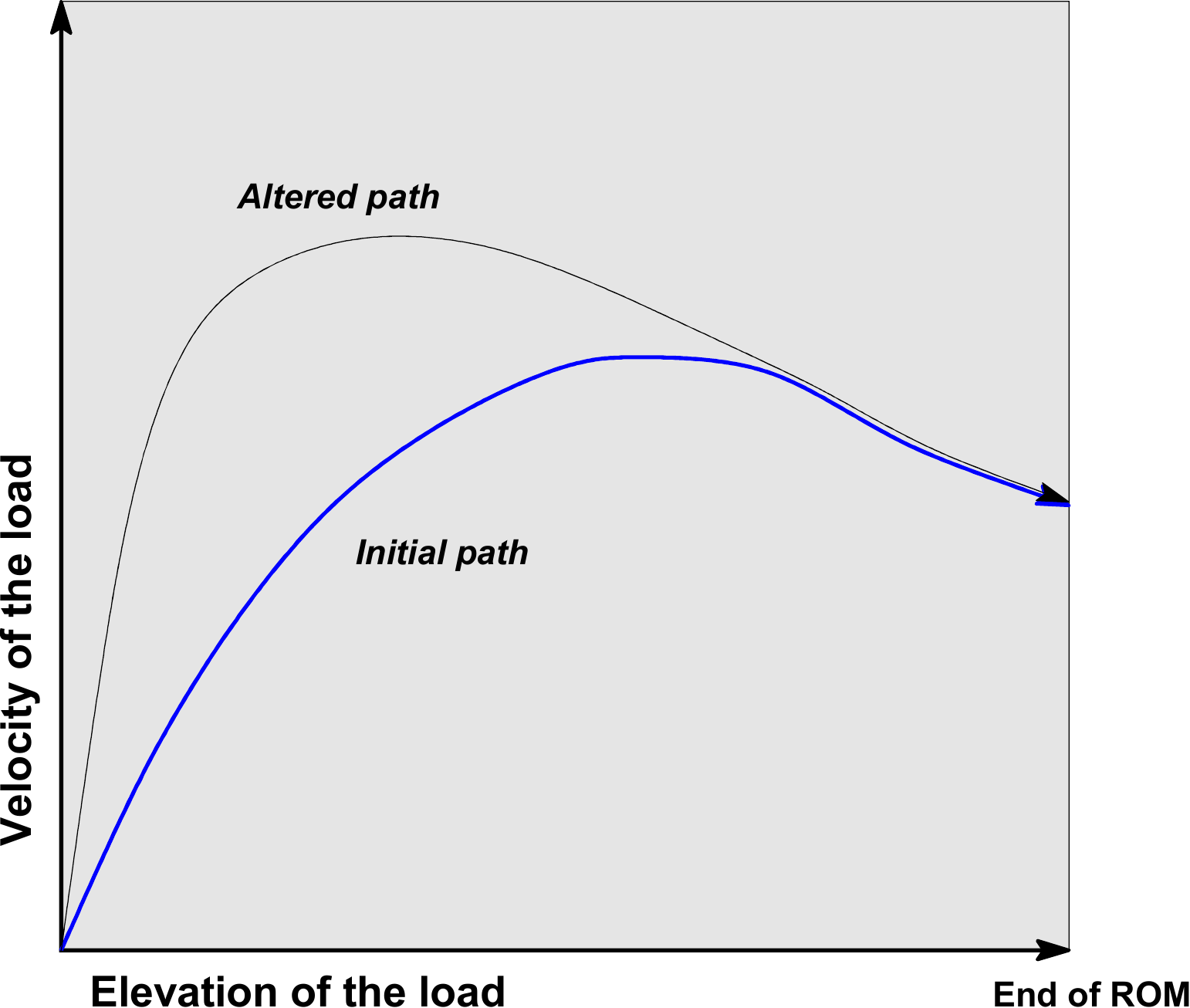}\\
    (c) && (d)\\
    &&\\[-5pt]
  \end{tabular}
  \caption{ Repetition profiles for the initial (thick blue line) and altered (thin black line) force production characteristics shown as paths in the capability plane described in~\cite{Aran2010-med}. Cases considered are: (a) overall performance is improved, (b) higher initial velocity is followed by inferior force production environment and the
  initial and altered paths intersection, (c) impossible inferior behaviour of the altered path, and (d) higher initial force production with no overall performance improvement. }
  \label{f:paths}
\end{figure}

Consider the capability plane paths $\dot{x}(x)$ and $\dot{x}_m(x)$ corresponding to respectively repetitions completed using the initial and modified force production characteristics (the reader should consult~\cite{Aran2010-med} for a detailed description of the methodology), the modification taking on the form of a localized increase in force production capability. If it holds that $\forall x.~\dot{x}_m(x) > \dot{x}(x)$, that is if for all points in the lift the velocity of the load becomes greater with the modification, the performance is clearly improved. This is illustrated in Figure~\ref{f:paths}(a). Thus, we need to consider the case when the two paths intersect for some $x_n$, as shown in Figure~\ref{f:paths}(b). For the overall performance to be affected negatively, after an intersecting point $x_n$ in the ROM\footnote{Note that we cannot yet assume that there is at most one intersecting point.}, the new path has to enter the region of the capability plane of lower velocities than $\dot{x}(x)$, i.e.\ it has to hold that $\dot{x}(x+dx) > \dot{x}_m(x+dx)$.

Next, notice that the fatigue accumulated up to the point $x_n$ is lower for the path $\dot{x}_m(x)$ than $\dot{x}(x)$. This is so because the velocity of the bar in the former case is consistently greater before the point of the ROM corresponding to $x_n$ and thus the time under tension shorter. Consequently, for the repetition performed using the modified force production characteristics, the force at $x_n$ when the velocity of the load is $\dot{x}_n \equiv \dot{x}(x_n)$ must be greater than that which allows the repetition to follow the original path. This means that the behaviour depicted in Figure~\ref{f:paths}(c) cannot occur. Therefore we can conclude that in the very worst case, the altered path $\dot{x}_m(x)$ will asymptote towards $\dot{x}(x)$ as the lift progresses, as shown in Figure~\ref{f:paths}(d). As such the performance is not worsened but merely not improved, as stated originally~\cite{Aran2011-med}.

\bibliography{./my_bibliography,./oa_physiology}

\begin{thebibliography}{5}
\providecommand{\natexlab}[1]{#1}
\providecommand{\url}[1]{\texttt{#1}}
\expandafter\ifx\csname urlstyle\endcsname\relax
  \providecommand{\doi}[1]{doi: #1}\else
  \providecommand{\doi}{doi: \begingroup \urlstyle{rm}\Url}\fi

\bibitem[Arandjelovi{\'c}(2010)]{Aran2010-med}
O.~Arandjelovi{\'c}.
\newblock A mathematical model of neuromuscular adaptation to resistance
  training and its application in a computer simulation of accommodating loads.
\newblock \emph{Eur J~Appl Physiol}, 110\penalty0 (3):\penalty0 523--538, 2010.

\bibitem[Arandjelovi{\'c}(2011)]{Aran2011-med}
O.~Arandjelovi{\'c}.
\newblock Optimal effort investment for overcoming the weakest point –- new
  insights from a computational model of neuromuscular adaptation.
\newblock \emph{Eur J~Appl Physiol}, 111\penalty0 (8):\penalty0 1715--1723,
  2011.

\bibitem[Arandjelovi{\'c}(2012)]{Aran2012-med}
O.~Arandjelovi{\'c}.
\newblock Common variants of the resistance mechanism in the {S}mith machine:
  analysis of mechanical loading characteristics and application to
  strength-oriented and hypertrophy-oriented training.
\newblock \emph{J~Stren Cond Res}, 26\penalty0 (2):\penalty0 350--363, 2012.

\bibitem[Arandjelovi{\'c}(2013{\natexlab{a}})]{Aran2013a-med}
O.~Arandjelovi{\'c}.
\newblock Does cheating pay: the role of externally supplied momentum on
  muscular force in resistance exercise.
\newblock \emph{Eur J~Appl Physiol}, 113\penalty0 (1):\penalty0 135--145,
  2013{\natexlab{a}}.

\bibitem[Arandjelovi{\'c}(2013{\natexlab{b}})]{Aran2013d}
O.~Arandjelovi{\'c}.
\newblock Computer simulation based parameter selection for resistance
  exercise.
\newblock \emph{Modelling and Simulation}, pages 24--33, 2013{\natexlab{b}}.

\end{thebibliography}
\bibliographystyle{plainnat}
\end{document}